\documentclass[11pt,preprint]{aastex}

\shorttitle{W51}
\shortauthors{Hodapp, Davis}

\begin{document}

\title{Molecular Hydrogen Outflows in W51}
\author{Klaus W. Hodapp}
\affil{
Institute for Astronomy, University of Hawaii,\\
640 N. Aohoku Place, Hilo, HI 96720
}
\email{hodapp@ifa.hawaii.edu}
\and
\author{Christopher J. Davis}
\affil{
Joint Astronomy Center,\\
660 N. Aohoku Place, Hilo, HI 96720
\email{c.davis@jach.hawaii.edu}
}

\begin{abstract} 
We present the results of a deep search for the molecular hydrogen shock
fronts associated with young stellar outflows in the giant molecular 
cloud and massive star forming region W51.
A total of 14 outflows were identified by comparing 
images in
the {\it H} and {\it K} bands and in a narrow-band filter 
centered on the H$_2$ 1--0 S(1) line at 2.122 $\mu$m. 
A few of the newly discovered outflows 
were subsequently imaged at higher
spatial resolution in the S(1) filter; 
one outflow was also imaged in the 1.644 $\mu$m emission
line of [FeII]. 
For two of the outflows, high-resolution echelle
spectroscopy in the H$_2$ 1--0 S(1) line was obtained using NIRSPEC at Keck.
For one outflow additional high resolution spectra were obtained
in the [FeII] line and in Br$\gamma$.
The largest and best studied outflow shock front shows a
remarkably broad [FeII] line, an unusual high-velocity 
component in Br$\gamma$, and
comparably narrow line widths in the
H$_2$ 1--0 S(1) line.
A scenario involving high-velocity shocks and UV excitation of
pre-shock material is used to explain these spectra.

\end{abstract}


\section{Introduction}

The high-mass star-forming region W51 was discovered as a region
of radio continuum emission by \citet{wes58} and as a massive
molecular cloud by \citet{wil71}. It is one of the 
most luminous star-forming regions in 
our Galaxy and may be the closest analog to the extremely luminous
star forming regions found in many other galaxies (e.g. 30 Doradus in
the LMC). W51 is far more
luminous than the well-studied relatively nearby Orion star forming region.
The mass of the W51 complex has been estimated to be about 10$^6$  $M_{\odot}$
\citep{car98}, based both on virial mass estimates
and $^{12}$CO intensities. Mass estimates of molecular clouds in
a complex environment obviously depend on where the boundaries are
set, and this estimate refers to a set of emission features of 
approximately spherical shape. W51 is in the top 1\% of all galactic
molecular clouds by size, and the top 5 - 10\% by mass.

W51 lies in the plane of the Galaxy at the substantial
distance of 7.0 $\pm$ 1.5 Kpc measured by \cite{gen81} 
through proper motion studies of the
W51 Main H$_2$O maser. W51 is therefore heavily
obscured by interstellar extinction (A$_V$ = 24 $\pm$ 3 mag, 
as measured by
\citet{gol94} on the basis of the color of the extended emission).
W51 is about 14 times farther away than Orion and the distance modulus
is almost 6 magnitudes larger
than for Orion, in addition to an extra 2 magnitudes of extinction in 
the K band, so that studies of
W51 are necessarily limited to very bright objects and rather large extended
features.

Numerous infrared sources were discovered in the radio 
continuum region W51.
One of these, W51 IRS 2 \citep{wyn74},
corresponding to the radio-continuum region W51e \citep{mar72}
is now known to be a young embedded star cluster larger than
the Orion-Trapezium cluster, as first suggested by \citet{gen82}. 
A detailed study of massive stars in W51 was done by \citet{oku00}.
They identified 4 subgroups within W51, roughly aligned parallel to
the Galactic plane, with average ages in those subgroups ranging from 
less than 1 Myr to 2.3 Myr.
Past studies of W51 have largely focussed on its global properties, e. g., the
total luminosity and mass, and have traced the effects of star formation
integrated over the lifetime of O-stars, by
studying radio continuum or Br$\gamma$ emission .

This paper, by contrast, seeks to obtain a snapshot of star formation 
activity in the most recent past. Stars in their main accretion phase
still have a substantial fraction of their final (ZAMS) mass
residing in a circumstellar disk. They are therefore
characterized by spectral energy distributions (SEDs) 
peaking at sub-mm wavelengths
\citep{and93} and are not directly detectable at near-infrared wavelengths. 
Fortunately, these class 0 
sources usually also have strong outflows that can be traced by the
shock-excited emission of H$_2$ resulting from the interaction
of the outflow with the ambient molecular cloud.
Examples of successful searches for outflows using narrow-band imaging in the 
H$_2$ 1--0 S(1) line
are, among many others, \citet{hod95}, one of the earlier
S(1) imaging surveys, and \citet{sta98}
who conducted a survey of large areas in the Orion GMC.
Very strong outflow activity and shock-excited molecular hydrogen emission
persists only through the class 0 and early class I phases, i.e. for
less than 10$^5$y. In contrast to the study of HII regions, 
H$_2$ outflows therefore
trace contemporary lower mass star formation.

We will discuss the technical aspects of the observations in chapter 2.
The results will be discussed in chapter 3, separated into discussions
of the overall distribution of outflow sources in chapter 3.1 and of
the morphology of individual outflows in chapter 3.2. Velocity resolved
spectroscopy of two outflows will be discussed in detail in chapters
3.3 and 3.4, and some results on the embedded IRS 2 cluster
will be presented in chapter 3.5.

\section{Observations and Data Reduction}

The data for this study of outflows in W51 were obtained in three
observing runs. The initial wide-field imaging data in {\it H}, 
H$_2$ 1-0 S(1) at 2.122 $\mu$m,
and {\it K} were obtained in 1997 and formed the basis for identifying 
previously unknown outflows. Some of the brightest outflows 
discovered in this survey were subsequently
imaged at higher spatial resolutions in S(1) and [FeII] in June 2000.
High resolution spectroscopy of two outflows was obtained at Keck
in May 2000 using NIRSPEC \citep{mcl98}.

\subsection{Wide-Field {\it H}, S(1), and {\it K}-band Images}

Our search for emission line objects in W51 is based on a wide-field
mosaic obtained in the broad band {\it H} and {\it K} filters and
in a narrow-band filter centered on
the H$_2$ 1--0 S(1) line at 2.122 $\mu$m.
The data were obtained 
with the infrared camera QUIRC \citep{hod96}
on 1997 August 16 -- 19 (UT)
at the UH 2.2 m telescope f/10 focus. 
The pixel scale was 0\farcs188 per pixel, 
resulting in a field of view of 193$\arcsec\times$193$\arcsec$.
To cover the large area of the W51 molecular cloud in a reasonable time,
we used 10 s integration time in the {\it H} and {\it K} filters and
20 s in the molecular hydrogen S(1) narrow-band filter.
The imaging data were obtained in a 10$\times$9 mosaicing pattern with
90$\arcsec$ spacing, so that every position on the sky is covered by at
least 4 individual exposures. The total integration time for each point
in the image is therefore 40 s in the broad-band filters and 80 s in the
S(1) narrow-band filter. 

The individual frames were flatfielded using domeflats and known bad
pixels were masked off. The individual reduced frames were assembled into
the large mosaic by aligning stars in the overlap region between of two
frames. The full {\it K}-band mosaic was assembled first, and then used
as a template to position the individual frames obtained in the other filters. 
This method ensures that all three colors are well registered relative to
each other, so that the three mosaic images can be combined into a false-color
image.

The three separate images in the three filters were then combined into
a false-color image, with {\it H} represented in blue, S(1) in green, and
{\it K} in red. Regions of S(1) line emission stand out as greenish features
in this color scheme. 
A compressed version of this false-color image can be downloaded from
\url{http://www.ifa.hawaii.edu/$\sim$hodapp/W51-sm-rgb.jpg}.
Visually identified potential shock front features were checked on the
individual {\it H}, S(1), and {\it K} mosaics and 
false detections due to residual images, ghost
images, or glare, as well as bad pixels or cosmic ray hits were discarded.

The extended emission permeating much of the survey area has been shown
by \citet{oku01} to be dominated by Br$\gamma$ and HeI emission, but to
also contain many emission lines of H$_2$, probably fluorescently excited
by the UV field from the ionizing stars in the HII regions. In identifying
shock-excited H$_2$ emission in the survey images, the distinction between
shock-excited features and features excited by fluorescence in the stellar
UV field can only be based on the strength of the H$_2$ emission, i.e. its contrast
relative to the broad-band images,  and its
morphology.
Objects identified as shock fronts in our survey have a strong contrast to the
broad-band filters, indicating that their emission in the {\it K} band is
dominated by H$_2$ emission. They are either marginally resolved small knots or
they clearly show the mophological features of bow shocks. In contrast,
extended, rather diffuse features showing some H$_2$ emission, but not being
dominated by it, that are also mostly located on the perimeter of the 
HII regions outlined in
broad-band {\it K} and the Br$\gamma$ image of \citet{oku00}, were not identified
as potential stellar outflow shock fronts. These latter features are most likely
excited by fluorescence in the UV radiation field generated by the ionizing
stars in the HII regions, as first suggested by \citet{oku01}.

The extent of the survey area and the location of detected S(1) emission
features fitting our search criteria is indicated in Fig. 1, 
the {\it K}-band image produced in our survey.
The individual emission knots, or systems of knots in cases of clearly
related (bipolar) outflow features, are shown in Fig. 2. For each
object, the name follows the IAU conventions and includes the J2000 coordinates.
Usually, the coordinates refer to the center of the extended emission region.
In highly structured, irregular emission knots, the coordinates refer to
the most prominent feature. For systems of emission features that 
morphologically appear to be part of one outflow, the coordinates refer to
one of the features, and additional features are referred to by additional,
self-explanatory designations (N, S, W, E, or C).

\subsection{High Resolution Emission Line Imaging}

To obtain images at higher spatial resolution of some of the
likely outflow shock fronts discovered in the wide-field survey, tip-tilt corrected
images were obtained on June 16 - 19, 2000 (UT). 
The tip-tilt corrected f/31 focus \citep{jim00} of the
UH 2.2m telescope was used where the QUIRC infrared camera has a pixel
scale of 0\farcs062 per pixel and a field of view of 63$\arcsec\times$63$\arcsec$.
Images were obtained in narrow-band [FeII] 1.644$\mu$m and 
H$_2$ 1--0 S(1) 2.122$\mu$m filters and the data were reduced in the same
way as described above for the wide-field images.
Under good seeing conditions, the tip-tilt corrected images at 2.122$\mu$m 
come close to the diffraction limit of the telescope and under more
typical conditions, 0.3$\arcsec$ FWHM is achieved.

\subsection{Infrared Spectroscopy}

High resolution echelle spectroscopy of two of the outflow sources in
W51 was obtained in the night of May 28, 2000 (UT) at the W.M. Keck Observatory,
using the NIRSPEC \citep{mcl98} spectrograph.
The weather in this night was non-photometric and unsuitable for
flux calibration of the spectra.

The spectra were obtained with two grating and filter settings:
The first setting was for 
observations in the {\it K} band, intended to cover several
of the H$_2$ shock-excited emission  lines but concentrating
on the 1-0 S(1) line at $\lambda_{vac}$=2.1218334$\mu$m \citep{bra82}. 
Observations in Br$\gamma$ were not part of the original observing
plan, since this line is rarely excited in H$_2$ outflows.
Fortunately, the Br$\gamma$ line at $\lambda_{vac} = 2.166167\mu$m was also
recorded in our {\it K}-band spectra, albeit near the edge of the
detector array so that wavelength calibration was difficult.
The other grating setting was for observations in
the {\it H} band, primarily to observe the [FeII] line 
at $\lambda_{vac}$=1.64400$\mu$m \citep{joh78} . 

Since the objects are extended, sky spectra were taken separately at
positions away from the visible shock fronts, but still within the
extended emission from the HII regions permeating 
much of the W51 star forming region.
The raw data frames were flatfielded with frames taken with continuum 
illumination of the spectrograph slit. We did not subtract dark current
frames from the individual exposures, but relied on the sky frames to
subtract dark current. 

Wavelength calibration was done using OH airglow lines recorded during
the on-object integrations. The line positions were measured on the flatfielded
frames before sky subtraction.
In the {\it H} band, 
we used the OH airglow lines (wavelengths from the UKIRT website)
at 1.63885$\mu$m, 1.64147$\mu$m,
1.64476$\mu$m, and 1.65023$\mu$m for wavelength calibration.
In the {\it K} band, OH lines at 2.11766$\mu$m, 2.12325$\mu$m, and 2.12497$\mu$m
were used for wavelength calibration near the H$_2$ 1--0 S(1) line.
As mentioned above, Br$\gamma$ was not part of the original observing
plan, but was nevertheless recorded near the edge of the detector array.
Wavelength calibration for Br$\gamma$ is based on extrapolation from
the two OH lines at 2.17111$\mu$m and 2.18022$\mu$m. Unfortunately, we
did not record any cataloged OH lines at wavelengths shorter than
Br$\gamma$ in the same echelle order, limiting the accuracy of the
wavelength calibration.
Only the [FeII] line, H$_2$ 1--0 S(1) and Br$\gamma$ were evaluated. 
Other transitions of
H$_2$ were detected in the spectrum, but not with sufficient signal
to noise to allow detailed analysis.
The slit width corresponds to a velocity resolution of 22 km/s.

In W51, \citet{oku01} have discussed near-infrared spectroscopy of the two
compact HII regions W51 IRS 2 East and West. They conclude that the faint
S(1) line emission in or around the HII region is excited by fluorescence rather
than shocks. This diffuse molecular hydrogen emission
is distinct from the shock-excited emission discussed here for the 
outflow objects. However, the presence of UV-excited fluorescent S(1)
line emission gives a background level of line emission that is clearly
detected in our spectra and not always fully removed by the sky-subtraction 
process due to its spatial variation.

\section{Results and Discussion}

\subsection{Distribution of outflow sources}

The locations of the shock-excited H$_2$ emission features are indicated
on the {\it K}-band image of the wide-field survey data (Fig. 1). 
One group of outflows is located in dense filaments near the main
HII regions associated with IRS 1 and IRS 2. The others are distributed
in almost a circular pattern at larger distances of about 6$\arcmin$ away from
this central group, on the perimeter of the HII region complex.
This distribution of outflow sources may be partly explained by
an observational selection effect, since our detection 
method is clearly not as sensitive in regions superposed on or lying behind
high levels of extended flux. 
More importantly, however, shock excited H$_2$ emission
cannot occur in fully ionized regions. 
In the likely case that the distribution of outflows is not entirely
dominated by selection effects, it
suggests that the class 0 and class I sources responsible for
these outflows may have formed as a result of secondary star formation
triggered by winds from the central HII regions and the cluster around
IRS2.

\subsection{Individual Molecular Outflows and Emission Knots}

The individual knots of H$_2$ emission found in this survey are
named following the IAU convention: The name indicates their 
association with the W51 complex and their discovery by emission of
the H$_2$ molecule and gives their J2000 coordinates with 1$\arcsec$ 
precision. In the following, we will discuss the morphology of each
candidate emission knot in detail.

\subsubsection{W51H$_2$ J192318.1+142738}

The H$_2$ emission W51H$_2$ J192318.1+142738 is a single, slightly
elongated patch of S(1) emission, about 1.5$\arcsec$ in NS extent (Fig. 2a).

\subsubsection{W51H$_2$ J192318.5+142656 and W51H$_2$ J192319.6+142654}

The emission knots W51H$_2$ J192318.5+142656 and
W51H$_2$ J192319.6+142654 are only separated by about $\approx$ 10$\arcsec$.
We show both of them in the high-resolution S(1) image Fig. 3
and discuss their relationship below.

Many H$_2$ outflows in nearby star-forming regions are associated with
class 0 sources that are not directly visible at near-infrared wavelengths.
Since mass accretion and the associated outflow power peak during the
class 0 phase, these extremely young sources often produce the strongest
S(1) shock fronts. It would therefore be entirely plausible that the two
shock fronts W51H$_2$ J192318.5+142656 and W51H$_2$ J192319.6+142654 
are part of the same bipolar outflow, even though there is no positive
detection of a central star that would drive the outflow. 

The arguments against this scenario are based on the morphology of
other extended emission features in the immediate vicinity of these two
brightest emission knots. 
The star $\approx$ 9$\arcsec$ north of
W51H$_2$ J192318.5+142656 is the reddest object within about $1\arcmin$
and shows some reflection nebulosity (based the neutral color 
of the extended emission
in the {\it H}, H$_2$ S(1) and {\it K} image) 
extending in the general direction of the 
W51H$_2$ J192318.5+142656 emission knot. 
It is therefore a plausible alternative scenario that
W51H$_2$ J192318.5+142656 originates in an outflow from that red
star (probably a class I source). 
The emission knot W51H$_2$ J192319.6+142654 shows extended emission
north and bending towards east from the main emission knot.
This emission is faintly indicated in Fig. 2a, showing that it is
mainly S(1) emission, and is clearly shown in the high-resolution
image in Fig. 3.
Based on these two morphological arguments, we list the two emission
knots as separate objects, probably belonging to different outflows.

\subsubsection{W51H$_2$ J192322.0+143333}

The S(1) line emission in W51H$_2$ J192322.0+143333 appears
marginally resolved in the high-resolution image in Fig. 4.
The only other faint extended feature in this image is the elongated
nebulosity 1$\arcsec$ east and 3.5$\arcsec$ south of W51H$_2$ J192322.0+143333
that points roughly to the latter. On morphological grounds, it could 
be nebulosity associated with
the driving source of W51H$_2$ J192322.0+143333, but is too faint to
be detected in the {\it H}, H$_2$ S(1), and {\it K} 
survey images, so its color is unknown.

\subsubsection{W51H$_2$ J192323.6+142515}

The pair of emission regions W51H$_2$ J192323.6+142515 S and N are clearly
part of the same bipolar outflow. The high-resolution S(1) image in Fig. 5
reveals a feature exactly in the middle of the bright N and S regions
that consists of an unresolved source and some nebulosity, separated
by about $1\arcsec$ along the axis between W51H$_2$ J192323.6+142515 S and
N, at coordinates 19:23:23.6 +14:25:15 (J2000). We label these features
as W51H$_2$ J192323.6+142515 C (center). Note that in this case
the two shock fronts of this outflow are listed as the north and south
features of an outflow listed under the coordinates of its central source. 
These central features are very red and are not visible at all in the
{\it H} band wide-field image. They are weakly indicated in the {\it K}-band
image. This is exactly the morphology seen in many closer bipolar
outflow sources, where two shock fronts are seen equidistant to
some scattered light near the central source. An example of this
is the outflow in L 1634 (Hodapp \& Ladd, 1995). For comparison, the extent of
the W51H$_2$ J192323.6+142515 outflow is $30\arcsec$ between the two
shock fronts, while the L 1634 outflow in Orion at a distance of 500 pc
is $6.2\arcmin$ long, nearly the same linear projected extent of $\approx$1 pc.

\subsubsection{W51H$_2$ J192325.9+143703 and W51H$_2$ J192327.9+143701}

Two faint emission knots in close vicinity were found at the northern 
edge of our survey field. The two panels in Fig. 2b showing 
W51H$_2$ J192325.9+143703 and W51H$_2$ J192327.9+143701 
actually overlap.
There is no morphological indication that the two emission knots are part
of the same outflow, so, conservatively, we list them separately, as
was discussed above. 

\subsubsection{W51H$_2$ J192335.0+143028 and W51H$_2$ J192336.6+143014}

The emission knot W51H$_2$ J192335.0+143028 is a bright, compact
knot of H$_2$ 1--0 S(1) line emission. Close to it is a second emission knot 
W51H$_2$ J192336.6+143014. It is noteworthy that these two emission
knots are collinear with one of the reddest stars detected in our
whole W51 survey area, labeled RS in Fig. 6,
at 19:23:32.7 +14:30:48 J2000.
The red star is undetected in our {\it H}-band image,
and appears much brighter in {\it K} (2.2$\mu$m) than in S(1) (2.12$\mu$m),
compared to other stars in the field.
Fig. 6 is a larger subframe of the survey images and
illustrates this collinear relationship between
these features. This alignment suggests that the red star may be the
driving source of a well-collimated jet of $\approx$ 2.2 pc projected
length that produces the two emission
knots. The degree of collimation of this jet would be quite remarkable,
but not unprecedented.

\subsubsection{W51H$_2$ J192338.3+143047}

The faint emission knot W51H$_2$ J192338.3+143047 has no morphological
features (Fig. 2c) that would help in identifying its driving source.

\subsubsection{W51H$_2$ J192339.7+143131}

The largest shock-front detected in S(1) is W51H$_2$ J192339.7+143131.
It appears
morphologically to be associated with a driving source in the
general direction of the IRS 2 young cluster to its south. 
A plausible candidate for the
driving source is the extremely red star at 19:23:39.8 +14:31:21
that lies close to the symmetry axis of the bow shock fronts.
More detailed images in the emission lines of
[FeII] and S(1) are shown in Fig. 7. The S(1) emission shows the
typical shape of multiple bow shocks, while [FeII] is generally
more concentrated
on the axis of the outflow, i.e. the areas of highest shock velocity and
excitation. It is noteworthy, however, that the S(1) emission exhibits
pronounced peaks near 
the apexes of two of the bow shocks, 
different from what is found in most other
S(1) bow shocks, e.g., \citep{ted99}. 
This object will be discussed in more detail below (3.4)
in the context of our spectroscopic data.

\subsubsection{W51H$_2$ J192345.5+143537}

The two 
emission knots found in
W51H$_2$ J192345.5+143537 are treated as part of one system, since their
close proximity makes a chance superposition of two unrelated shock
fronts very unlikely. The high-resolution
image in Fig. 8 shows one knot clearly resolved, the other marginally
resolved, but no further conclusions about the location of the driving
source can be derived from these images.

\subsubsection{W51H$_2$ J192347.2+142944}

The star labeled as W51H$_2$ J192347.2+142944 stands out due to its 
strong flux in the S(1) filter. While the star appears slightly more
extended than other stars on the S(1) frame and more extended than
its {\it K}-band image, the distribution of S(1) emission around the star
cannot be mapped out. We conclude that the molecular emission in this
region originates in the immediate vicinity of the star.

\subsubsection{W51H$_2$ J192403.3+143255}

The emission knot W51H$_2$ J192403.3+143255 is extended, but small.
No obvious association with other emission knots or red stars was
found that might help in identifying the driving source.

\subsection{Velocity-resolved spectroscopy of W51H$_2$ J192323.6+142515 S}

A high spatial resolution image of the two H$_2$ emission features 
W51H$_2$ J192323.6+142515 N and S is shown in
Fig.\,5 and was discussed in section 3.2.4. 
A high-resolution H$_2$ spectrum was obtained towards only the
southern component at the slit position indicated in Fig.\,5
and is shown in Fig.\,9. 

In knot {\it S-A}, farthest from the likely driving source, broad,
double-peaked H$_2$ profiles peaking at $V_{\rm LSR} \sim 60$\,km
s$^{-1}$ (just blue of the cloud systemic velocity of 70\,km s$^{-1}$) and 
$V_{\rm LSR} \sim 0$\,km s$^{-1}$
(blue-shifted by $\sim 70$\,km s$^{-1}$) are observed.  
This  velocity-split emission corresponds to two spatially separate emission
knots in Fig.\,5. Note however that the separation of the two velocity
components is much larger than the spectral resolution of 22 km/s,
so the velocity split is real and not just a projection of two spatially
separate knots in the slit.
Behind (north of) the shock front {\it S-A} (offset 1\arcsec\ in
Fig.\,9) the H$_2$ velocities converge to an intermediate, blueshifted
velocity $V_{\rm LSR} \sim 20$\,km s$^{-1}$, spatially coinciding with
a single emission knot centered on the slit. 

The fainter knot {\it S-B} (not covered well by the slit)
also shows a double-peaked H$_2$ profile with
peaks at $V_{\rm LSR} \sim 40$\,km s$^{-1}$ and 
$V_{\rm LSR} \sim -20$\,km s$^{-1}$, i.e. about 20 km s$^{-1}$ more
blueshifted than knot S-C.
Further back towards the driving source, shock front {\it S-C} 
is much more diffuse than
{\it S-A} and {\it S-B}, and fainter. Fig. 9 shows a broad line centered at
$V_{\rm LSR} \sim 30$\,km s$^{-1}$, without a clear indication of
a separation into two peaks.

The observed double-peaked H$_2$ profiles, with peak-to-peak
separations of $\sim 60$\,km s$^{-1}$, are predicted by numerical models
\citep{vol99} and are implied by the analytical 
\citep{har87} J-type bow shock models of
atomic line emission.  
The most extended, double-peaked profiles will be expected near the
front of the bow shock; narrow, low-velocity peaks will instead be
associated with the oblique wings.
Indeed, the range of H$_2$ velocities observed
in W51H$_2$ J192323.6+142515 S can be explained on purely geometrical grounds,
if the flow is inclined towards the observer at an angle (with respect
to the line of sight) of $\phi = 40^{\circ} - 70^{\circ}$.
By comparison, a bow shock moving in the plane of the sky ($\phi =
90^{\circ}$) would produce two (blended) peaks blue- and red-shifted
by the same amount, while a bow viewed head-on would produce only
one, blue-shifted component (see for example Plots IV and I in the
appendix in \citet{dav01}.  

Overall, the spectral features seen in W51H$_2$ J192323.6+142515 S are
quite similar to those found in HH 111 by \citet{dav01}. They can
be explained by the straightforward geometric fact that the H$_2$ emission
arises in the oblique shocks in the wings of the bow shocks, where
shock velocities are low despite a high velocity of the jet relative
to the ambient cloud medium. The shock velocities seen 
in W51H$_2$ J192323.6+142515 S
are just above the dissociation limit for pure J shocks, which suggests
a certain degree of magnetic cushioning to avoid rapid dissociation
of the H$_2$ molecule \citep{smi94} .

\subsection{Velocity-resolved spectroscopy of W51H$_2$ J192339.7+143131}

\subsubsection{Morphology and slit orientation}

The bow shocks in W51H$_2$ J192339.7+143131 indicate a
much faster and powerful outflow than the outflow discussed above.
High-spatial resolution images of W51H$_2$ J192339.7+143131 
in [FeII] and H$_2$ 1--0 S(1) are shown in
Fig. 7.  This object is situated 23$\arcsec$ north of the
IRS 2 cluster.

In H$_2$,
W51H$_2$ J192339.7+143131 resembles a sequence of at least three ``nested''
bow shocks separated by 2\arcsec --3\arcsec,
labeled {\it A}, {\it B} and {\it C} in Fig.7. 
A very similar arrangement of nested bow shocks, with
the larger ones being found in the wakes (i.e. towards the driving source) of
smaller bow shocks, is observed in the low-mass L\,1634
outflow \citep{hod95}. The bow shock nature of
W51H$_2$ J192339.7+143131 is confirmed by high-resolution spectroscopy in the
H$_2$ 1--0 S(1) and [FeII] lines (Fig. 10).

The [FeII] emission in W51H$_2$ J192339.7+143131 (Fig. 7, left panel)
appears to be generally more closely
confined to the north-south outflow axis than the H$_2$ emission.
This is not unexpected since the [FeII] traces the
higher-excitation bow shock caps while the H$_2$ is excited in the
oblique, lower-excitation bow-shock wings \citep{dav00,lor02}.
However, contrary to this generalized statement, we find strong
H$_2$ emission from near the apexes of the bow shocks, which we 
believe to result from fluorescence, as will be discussed below.

We present [FeII], H$_2$, and Br$\gamma$ (Fig. 10) spectra
observed through the
center of W51H$_2$ J192339.7+143131.  
The slit was aligned to cover the brightest H$_2$ 1--0 S(1) emission knots
({\it B} and {\it C}) and is probably closely aligned
with the outflow axis, if our identification of the driving source
or at least the location of the driving source in the general area
of IRS 2 is indeed correct.
Millimeter-wave mapping of the cloud
structure in W51A suggests that the LSR systemic velocity of the
IRS\,2 region is $\sim 61$\,km s$^{-1}$ \citep{car98},
although variations in the systemic velocities of cloud cores across the
region are evident in the molecular cloud maps.
Spatially
extended S(1) emission around IRS 2, probably UV-excited in the radiation
field of the 0 stars in IRS 2, and largely
subtracted out from our sky-subtracted spectral images, is centered around
70 km/s. All three spectra were individually wavelength calibrated using
OH airglow lines recorded on the object frames. The wavelength calibration
of the Br$\gamma$ line is relatively poor, since Br$\gamma$ was not part
of the original observing plan and was only evaluated when interesting
structure was unexpectedly found. As a consequence, Br$\gamma$ was recorded
near the edge of the detector array where optical distortions could be
significant, and the wavelength calibration could only be done by extrapolation
from OH-lines, not by interpolation, since there were no OH lines recorded
shortward of 2.166 $\mu$m in the echelle order used. The small differences
in the velocity of the narrow components 
of S(1) and Br$\gamma$ emission are probably
due to these calibration problems and we do not believe they are
significant. 
We adopt a systemic cloud velocity of $\sim 70$\,km s$^{-1}$
for IRS2 and W51H$_2$ J192339.7+143131.  

The position-velocity images in
Figs. 10 show very distinct features in the three emission lines
studied here.
The H$_2$ 1--0 S(1) emission is largely confined to a narrow velocity range
around the systemic velocity. 
The velocity structure seen in H$_2$ 1--0 S(1) is, however,
correlated to the high-velocity features seen in [FeII] and Br$\gamma$,
in the sense that knot {\it B} shows a slight blueshift 
and knot {\it C} a slight
redshift. Only in the faint knot {\it A}, north of the main shock fronts,
does the S(1) line split, producing a weak blueshifted component, as expected
for the spectrum of a non-dissociative bow shock.

The [FeII] line is a more robust tracer of shocked gas than the
rather fragile H$_2$ molecule and more closely traces the full velocity field
in the shock fronts, which in this case extends over a range of 300 kms$^{-1}$. 
The detection of velocity features related to the shock fronts
in Br$\gamma$ was a surprise. We had, of course, expected copious
Br$\gamma$ emission from the nearby IRS2 HII regions, either
projected as foreground or background emission or scattered into
the line of sight. However, the Br$\gamma$ line in W51H$_2$ J192339.7+143131
shows features clearly correlated to those seen in the other two
lines studied here. These features are clearly related to the
outflow shock fronts and make W51H$_2$ J192339.7+143131 a rare case
of outflow shock fronts with detectable Br$\gamma$ emission.
We will now discuss the individual emission lines in order of
wavelength.

\subsubsection{The [FeII] Line}

The [FeII] emission is concentrated in the region between the
H$_2$ knots {\it B} and {\it C} and is 
less spatially extended perpendicular to the
jet axis than the S(1) emission (Fig. 7). This is consistent with the
fact that [FeII] emission should be concentrated in the high
excitation regions directly at the head of the bow shock, a region where
S(1) emission is suppressed by dissociation of the H$_2$ molecule.

At the position of knot {\it C} and up to 1.5\arcsec north of it,
the [FeII] emission is strongly redshifted, and very broad,
extending from $V_{\rm LSR} \sim 20$\,km s$^{-1}$ to
$V_{\rm LSR} \sim 200$\,km s$^{-1}$. Further north, up to the
position of knot {\it B}, broad emission is observed
with velocities ranging from blueshifted $V_{\rm LSR} \sim -60$\,km s$^{-1}$
to redshifted $V_{\rm LSR} \sim 120$\,km s$^{-1}$.
Overplotted on the spectral image in Fig. 10 is the normalized spectrum
integrated over the relevant parts of the slit, for comparison
with the model presented by \citet{har87}. The broad,
only slightly asymmetric profile measured in W51H$_2$ J192339.7+143131
closely matches the theoretical profile calculated for H$\alpha$ emission of 
their 200 km/s jet model
inclined by 60$^{\circ}$ against the line of sight towards the
observer (their Fig. 3c). 
In our case, the total velocity spread is larger, about
300 km/s and the redshifted component is somewhat stronger and
more extended than the model would predict, but the overall
agreement to the idealized model is remarkably good.

The [FeII] spectrum shows a much wider spread of velocities than
the S(1) spectra at the same slit position. Our S(1) and [FeII] images
suggest a spatial anticorrelation between the two emission lines. S(1)
is primarily emitted in front and behind the main shock outlined
in [FeII], since the high temperatures in this 300 km/s shock
dissociate the H$_2$ molecule.

In addition to the good agreement of the integrated [FeII]
spectrum with the models by \citet{har87}, the details of the [FeII]
spectrum, i.e., the wide velocity range
observed and the prominent redshifted emission
can be explained by a fully developed hydrodynamical model.

The hydrodynamical models of
\citet{vol99} discuss the gas motions near the apex of
a jet in detail. Their Fig. 13 clearly shows that, in the reference
frame of the jet, gas is being pushed sideways at the apex of the jet
and streams backward in the wings of the bow shock.
The same models also show that the
velocity field in the bow of a shock can be complex and that
knots of peculiar velocities exist. 
In the case of W51H$_2$ J192339.7+143131 where the predominantly
blueshifted and redshifted components of [FeII] emission
are spatially separated,
we have to postulate that the emission arises from distinct
emission knots near the apex of the bow, the redshifted emission
being on the far side of the jet and therefore receding relative
to the jet, while the blueshifted emission is from the near side
of the bow, moving towards the observer in excess of the jet velocity.

\subsubsection{The H$_2$ 1--0 S(1) Line}

The H$_2$ 1--0 S(1) line in W51H$_2$ J192339.7+143131
is strikingly narrow in an object that
shows such substantial line width in [FeII]. 
Only the northernmost and faintest of the emission knots identified in 
Fig. 7 (knot {\it A}) shows the split line profile expected 
from a bow shock seen from the side. 
The presence of [FeII] emission supports J-type shock excitation,
i.e., no significant cushioning by magnetic fields.
In such shocks, H$_2$ will be dissociated at relatively low
shock velocities, of the order of 20-25 kms$^{-1}$.
The centers of the two velocity components in
knot {\it A} are separated by about 50 kms$^{-1}$, which is roughly 
double the
dissociation shock velocity limit for H$_2$. 
The other two knots ({\it B} and {\it C}) do not
show a splitting of the lines, but show slightly broadened
spectral features spreading over a maximum of 50 kms$^{-1}$, i.e.
within the dissociation limit. The velocity structure is related
to the velocity field seen in [FeII] emission in the sense that
knot {\it B} exhibits a slight blueshift while 
knot {\it C} is slightly redshifted.
H$_2$ will survive and be collisionally excited into emission
only in the oblique wings of the bow shocks, exhibiting narrow
line profiles. An example of this are the bullets in Orion \citep{ted99}
where the tips of the bows are traced in [FeII] and the wings
in H$_2$ S(1). 
However, in our images the H$_2$ emission appears brightest on-axis, 
in the two compact knots {\it B} and {\it C}
near the apex of the bow shocks, where molecules will be collisionally
dissociated. 
We therefore postulate that the strong, spatially concentrated H$_2$ emission
near the apex of the bow
with narrow line width represents fluorescent emission from 
quiescent molecular material in front of the bow shock that is excited
by the UV radiation field generated by the high temperature in
the shock front. The presence of Br$\gamma$ emission in the
shock front is proof that temperatures high enough to
dissociate H$_2$ and ionize H exist locally in the shock front.
Fluorescent emission from H$_2$ in the UV has been
found in IUE spectra of several HH objects by \citet{boh91}.
In an object similar to W51H$_2$ J192339.7+143131, the well studied
HH 7, the higher excitation lines of H$_2$ have 
been explained
by fluorescence of H$_2$ in the presence of a strong UV field 
generated in the shock fronts \citep{fer95}.

\subsubsection{The Br$\gamma$ Line}

Emission in the Br$\gamma$ line is very rarely seen in the shock
fronts of stellar outflows, even though H$\alpha$ is
commonly observed in low-extinction Herbig-Haro jets and
shock fronts (see \citet{rei01} for a recent review).
Br$\alpha$ has recently been observed \citep{ful01} in an outflow
from a high-mass star.

Emission of Br$\gamma$ is, of course, expected near the HII region
associated with IRS2 and could arise in the foreground or background
of the outflow shock fronts of W51H$_2$ J192339.7+143131, or be
scattered into that line of sight. Remarkably, however, the 
Br$\gamma$ emission in W51H$_2$ J192339.7+143131 shows features
similar to those seen in [FeII] and H$_2$, demonstrating that 
a significant fraction of the Br$\gamma$ flux originates in
the outflow shock fronts themselves. 

The brightest peak in the Br$\gamma$ position-velocity image is
just north of knot {\it B}. 
The narrow-line Br$\gamma$ intensity south of knot {\it B} drops to
55\% of its value (per pixel) just north of this knot. This indicates
that $\approx$45\% of this narrow-line flux must originate
close to the shock front, while $\approx$55\% may be foreground or
background flux from elsewhere along the line of sight.
The Br$\gamma$ emission in knot {\it C}
is more spatially extended than the H$_2$ emission, but they are clearly
related. 

At very low but clearly significant flux levels,
spectrally very broad Br$\gamma$ emission essentially duplicates
the features seen in [FeII]: strong blueshift in knot {\it B} and
redshift in knot {\it C}, with about the same total velocity extent.
This component of the Br$\gamma$ emission is clearly related to
fast moving gas in the bow shock front. 
The broad high-velocity Br$\gamma$ emission is probably collisionally
excited in the rapid, high-excitation shock fronts, similar
to the [FeII] emission.

As explained above, absolute velocity
calibration of the Br$\gamma$ is rather uncertain due observational
limitations. We therefore do not ascribe any significance to the
absolute velocity difference between the H$_2$ 1--0 S(1) 
and Br$\gamma$ lines in Fig. 10.

\subsubsection{Synopsis of W51H$_2$ J192339.7+143131}

Of the three emission lines discussed here, [FeII] is a pure
and robust tracer of shock-excited gas. It shows a spread of
velocities over a range of 300 kms$^{-1}$ and its integrated
line profile matches simple models of bow shock emission very 
well. The Br$\gamma$ line traces both the high-velocity gas,
either by direct shock excitation or UV excitation from the shock, 
but also traces
the low velocity components, by UV excitation from the shock
in addition to fore and background emission. The H$_2$ emission
traces only low-velocity gas, since H$_2$ dissociates in high
velocity shocks. The strong H$_2$ emission from the tips of the
bow shocks can best be explained by fluorescent excitation of
H$_2$ molecules in the radiation field of the shock, just prior
to their being dissociated when hit by the shock directly. 

An outflow system appearing superficially similar to W51H$_2$ J192339.7+143131
are the ``bullets'' observed to emerge from a source in the
Orion-Trapezium cluster. A detailed study of this system, including
high-resolution spectroscopy in the [FeII] and H$_2$ S(1) lines
has recently been presented by \citet{ted99}.
The shock fronts in W51H$_2$ J192339.7+143131 appear to have a
more organized large scale shape than the system of small
''bullets'' in Orion, but some small-scale clumpy structure is
clearly present in W51H$_2$ J192339.7+143131.
The Orion ''bullets'' do not show the very broad [FeII] profiles
seen in W51H$_2$ J192339.7+143131 and they do not exhibit the
strong H$_2$ emission at the apex of the bow shock. Rather, they
show H$_2$ only in the wings of the bow shocks, as expected from
pure shock excitation.

Observations in [FeII] of the jets emanating from L1551 IRS5 have
recently been reported by \citet{pyo02}. Their spectra along the
jet axis show a similarly broad velocity profile of the [FeII] line
to the one reported here. In contrast to our observations, theirs
concentrated on the jet close to its source of origin and did not
include well developed bow shocks.

\subsection{The young stellar cluster around IRS2}

The narrow-band H$_2$ 1--0 S(1) line image was also used to study the young
cluster surrounding \mbox{IRS 2} in more detail. 
This image reaches about the same
limiting magnitude and has better spatial resolution than the {\it K}-band
image obtained of the wider W51 region, and is therefore best suited
for a rough count of the stars contained in this cluster. The S(1) line
image was photometrically calibrated using the standard star FS 27
\citep{haw01}
so that approximate {\it K}-band magnitudes could be obtained. Down to a 
limiting magnitude of {\it K}=15.3, we count 60 stars within a projected
circle of 1 pc diameter.
\citet{oku00} have already
pointed out that one foreground star is visible in front of the 
W51 IRS 2 cluster, but clearly most of the stars counted around IRS 2
are physically close to this embedded source. 
The image (Fig. 11) also shows areas very near IRS 2 that appear as dark
patches against the extended flux of the HII region and with virtually no
stars, indicating opaque extinction. Roughly half of the 1 pc
diameter circle is covered by such extinction. 

The distance of 7Kpc to W51, compared to 0.5 Kpc to the 
Trapezium cluster, makes stars
appear 5.73 magnitudes fainter. \citet{oku00} found a strong
peak at A$_V$=25 mag in their extinction histogram of "region 3" that
contains the young embedded cluster associated with IRS2. We take
A$_V$=25 here as an estimate of the extinction to 
those stars in the cluster that
are not completely obscured, obviously an extreme simplification of
the actual situation. Similar A$_V$ values were found for many of
the stars in the immediate vicinity of IRS2 by \citet{gol94}.
While infrared excess may be partly responsible for the red colors leading
to these extinction estimates, we assume here that stars in the W51 IRS2 
cluster suffer about A$_K$=2 mag more extinction than the low-extinction
stars seen in the Orion Trapezium \citep{her86}. This statement obviously excludes 
the high extinction objects in the BN/KL region. 
Combining the effects of distance modulus and extinction, stars in
the W51 IRS2 cluster appear about 7.7 mag fainter than they would in
the Orion Trapezium cluster. Our limiting {\it K}-band magnitude of 15.3 
in the high-resolution image in Fig. 12 corresponds
to a magnitude of 7.6 in Orion.
On the 2MASS {\it K}-band frame we count 15 stars brighter 
than {\it K}=7.6 mag in
a circle of projected diameter of 1 pc, using photometry in the
Trapezium cluster from \citet{mcc94}. 

In comparing the young stellar clusters associated with the Orion Trapezium
and W51 IRS 2, we can reach a few tentative conclusions, despite the
uncertainties in the estimates for extinction and the incompleteness of
the star count due to opaque areas obscuring about half the cluster,
if one assumes a roughly spherical intrinsic distribution of the stars.
The star count in the W51 IRS2 cluster appears between 4 to 8 times
higher than in the Trapezium cluster. This is roughly consistent with
the count of O stars by \citet{gol94} (4 in Trapezium vs. probably 9 in
W 51 IRS 2). 

\section{Conclusions}

Our {\it H}, H$_2$ 1--0 S(1), and {\it K}-band survey of the 
W51 GMC resulted in the
discovery of 14 H$_2$ shock fronts associated with stellar outflows.
The outflows are found in dense molecular filaments near the
central HII regions and the young embedded cluster associated
with IRS 2, and near the perimeter of the molecular cloud complex,
away from the HII regions. We speculate that the outflows near the
perimeter represent a secondary, triggered phase of star formation
in W51. 

Detailed, high-resolution images were used to identify plausible
candidates for the driving sources of some of the outflows.

For two of the outflow shock fronts, high resolution spectroscopy
was obtained. Of particular interest are the shock fronts 
W51H$_2$ J192339.7+14313. The very broad [FeII] line seen in this
shock front is in good agreement with models developed by
\citet{har87} for other atomic lines in shock fronts.
Refinements of these models based on the hydrodynamic models
of \citet{vol99} lead to a satisfactory agreement with
the observations, even though the large observed line width and
the large redshifted component remain quite remarkable.
We also report the very rare detection of high-velocity Br$\gamma$
emission from this outflow, with a high-velocity components essentially
matching the [FeII] profiles.
The strong, narrow H$_2$ 1--0 S(1) emission found near the apex of the
shock fronts strongly suggests fluorescence as the excitation
mechanism for H$_2$ at this location.
The outflow
W51H$_2$ J192339.7+14313 is driven by a source associated with the
IRS 2 young embedded cluster. Based on star counts and rough
extinction estimates, we conclude that this cluster is
more massive, richer in stars and richer in high-mass stars than
the Orion-Trapezium cluster, in agreement with previous results
by others.

\acknowledgements
The imaging data presented here were obtained at the University of
Hawaii 2.2m telescope.
The high-resolution spectroscopy presented here is based on data
obtained at the W. M. Keck
Observatory, which is operated as a scientific partnership among the
California Institute of Technology, the University of California and the
National Aeronautics and Space Administration. The Observatory was made
possible by the generous financial support of the W. M. Keck Foundation.

The authors wish to extend special thanks to those of Hawaiian ancestry 
on whose sacred mountain we are privileged to be guests.

\clearpage

\begin{center}
{\bf Figure Captions}
\end{center}

\figcaption{Wide-field image of W51 taken in the {\it K}-band using QUIRC
at the UH 2.2 m telescope in 1997. The circles indicate the positions where
H$_2$ line emission features were found. \label{fig1}}

\figcaption{
{\it H}, S(1), and {\it K}-band images of the newly discovered outflows in W51.
}




\figcaption{
High-resolution tip-tilt corrected image 
at 2.12 $\mu$m (H$_2$ 1--0 S(1)), showing
shock-excited emission in the W51H$_2$ J192318.5+142656
and W51H$_2$ J192319.6+142654 shock fronts.
The images were taken in June of 2000 using QUIRC at the f/31 focus
of the UH 2.2m telescope.
}

\figcaption{
High-resolution tip-tilt corrected image 
at 2.12 $\mu$m (H$_2$ 1--0 S(1)), showing
shock-excited emission in the 
W51H$_2$ J192322+143333 emission knot.
The images were taken in June of 2000 using QUIRC at the f/31 focus
of the UH 2.2m telescope.
}

\figcaption{
High-resolution tip-tilt corrected images 
at 2.12 $\mu$m (H$_2$ 1--0 S(1)), showing
shock-excited emission in the W51H$_2$ J192323.6+142515 North and South
shock fronts, and the faint emission features near the possible central
source of this bipolar outflow.
Overlayed on the image of W51H$_2$ J192323.6+142515 S is the slit 
position for the NIRSPEC high-resolution spectrum.
The images were taken in June of 2000 using QUIRC at the f/31 focus
of the UH 2.2m telescope.
}

\figcaption{
Wide field H, S(1) and K-band images of
the emission knots
W51H$_2$ J192335.0+143028 and W51H$_2$ J192336.6+143014.
All three wavelengths of the wide-field survey images
are shown to illustrate the extreme red color of the
star (labeled RS) at 19:23:32.7 +14:30:48 J2000,
that lies on the line connecting the two emission
knots, and that is therefore a plausible candidate
for the driving source.
}

\figcaption{
High-resolution tip-tilt corrected images at 1.64 $\mu$m ([FeII]), 
and 2.12 $\mu$m (H$_2$ 1--0 S(1)), showing
shock-excited emission in W51H$_2$ J192339.7+143131 . 
Overlayed on the images is the slit position for the NIRSPEC high-resolution
spectra.
The images were taken in June of 2000 using QUIRC at the f/31 focus
of the UH 2.2m telescope.
}

\figcaption{
High-resolution tip-tilt corrected image 
at 2.12 $\mu$m (H$_2$ 1--0 S(1)), showing
shock-excited emission in the W51H$_2$ J192345.5+143537 shock front.
The images were taken in June of 2000 using QUIRC at the f/31 focus
of the UH 2.2m telescope.
}

\figcaption{
High-resolution spectrum of W51H$_2$ J192323.3+142515 S in the
2.122 $\mu$m H$_2$ 1--0 S(1) line.  The slit position is as indicated in
Fig. 5. The labels {\it S-A}, {\it S-B}, and {\it S-C} refer to the 
emission knots identified in Fig. 5.
Overplotted on the spectral image are the extracted, normalized spectra
at four different locations in the shock fronts. The horizontal marks
indicate the zero-intensity level, the vertical marks indicate the
spatial region along the slit which the spectrum was integrated over.
}

\figcaption{
High-resolution spectra of W51H$_2$ J192339.7+143131 in the
1.644 $\mu$m line of [FeII], the 2.118 $\mu$m 1-0 S(1) line of H$_2$, and
the 2.166 $\mu$m Br$\gamma$ line of H. The slit position is as indicated in
Fig. 7. A normalized spectrum of the [FeII] emission, integrated along
the slit, is overplotted. Each spectrum is given in two different linear
stretches, to show faint emission features. 
}

\figcaption{
High-resolution tip-tilt corrected images at 
2.12 $\mu$m (H$_2$ 1--0 S(1)), showing the young embedded stellar
cluster associated with W51 IRS 2.
The image was taken in June of 2000 using QUIRC at the f/31 focus
of the UH 2.2m telescope.
}

\end{document}